\documentclass[12pt]{article}
\usepackage{amssymb,amscd,array}
\catcode `\@=11
\@addtoreset{equation}{section}

\def\d{\partial}
\def\dh{\mathop{\vphantom{\odot}\hbox{$\partial$}}}
\def\dl{\dh^\leftrightarrow}

\newcommand{\be}{\begin{equation}}
\newcommand{\ee}{\end{equation}}
\newcommand{\bea}{\begin{eqnarray}}
\newcommand{\eea}{\end{eqnarray}}
\begin{document}
\begin{titlepage}

\begin{center}
{\bf \Large{Non-trivial 2+1-dimensional Gravity\\}}
\end{center}
\vskip 1.0truecm
\centerline{D. R. Grigore, 
\footnote{e-mail: grigore@theory.nipne.ro}}
\vskip5mm
\centerline{Department of Theoretical Physics, Institute for Physics and Nuclear
Engineering ``Horia Hulubei",}
\centerline{Institute of Atomic Physics}
\centerline{Bucharest-M\u agurele, P. O. Box MG 6, ROM\^ANIA}
\vskip 1cm
\centerline{G. Scharf
\footnote{e-mail: scharf@physik.unizh.ch}}
\vskip5mm
\centerline{Institut f\"ur Theoretische Physik, Universit\"at Z\"urich,} 
\centerline{Winterthurerstr. 190 , CH-8057 Z\"urich, SWITZERLAND}

\vskip 2cm
\bigskip \nopagebreak
\begin{abstract}
\noindent
We analyze 2+1-dimensional gravity in the framework of quantum gauge theory.
We find that Einstein gravity has a trivial physical subspace which reflects
the fact that the classical solution in empty space is flat.
Therefore we study massive gravity which is not trivial. In the limit of
vanishing graviton mass we obtain a non-trivial massless theory different
from Einstein gravity. We derive the interaction from descent equations and
obtain the cosmological topologically massive gravity. However, in addition to
Einstein and Chern-Simons coupling we need coupling to fermionic ghost and
anti-ghost fields and to a vector-graviton field with the same mass as the graviton.

\end{abstract}
\end{titlepage}

\section{Introduction}

It is well known that 3-dimensional Einstein gravity is dynamically trivial in the sense
that any classical solution in empty space is flat \cite{1}-\cite{3} (and references given there).
For this reason one has to consider modifications of Einstein's theory, in particular
massive gravity \cite{4}-\cite{18} in order to have propagating gravitational waves. Most of these
studies have been done in the framework of classical Lagrangian field theory.

We advocate a different approach to gravity which has been successfully applied in the
3+1-dimensional case \cite{8}-\cite{10}. We do not start from a classical Lagrangian. Instead we choose a
collection of free quantum fields on Minkowski space which are the asymptotic fields of
a S-matrix. Some of these fields are gauge fields which have a non-trivial gauge variation
involving fermionic ghost fields. The gauge variation 
$
d_Q=[Q,\cdot]
$ defines the physical subspace
of the theory as 
\be
{\cal H}_{\rm phys}={\rm Ker}Q/{\rm Ran}Q.
\ee
The coupling 
$
T(x)
$ 
is a Wick polynomial of the free fields which is a solution of the
gauge invariance condition
\be
d_Q T=\d_\alpha T^\alpha(x),
\label{gauge-inv}
\ee
and some generalization of it where 
$
T^\alpha(x)
$ 
is another Wick polynomial. This condition
is necessary in order to have unitarity of the S-matrix on the physical subspace. In 3+1
dimensions the method works equally well in the massless and massive case. The difference is 
only that the gauge structure in the massive case is more rich. It requires the so-called vector
graviton field 
$
v_\mu
$ 
with the same mass $m$ as the graviton, so that the resulting theory
is a vector-tensor theory. The interesting point is that in the limit 
$
m\to 0
$ 
one does not get Einstein's theory. The now massless vector-graviton field does not decouple from the
symmetric tensor field 
$
h_{\mu\nu}
$ 
so that we get an alternative (massless) gravity theory.

It is the purpose of this paper to Analise 2+1-dimensional gravity in exactly the same way.
In the next section we introduce the various free quantum fields and their gauge structure.
Then we determine the physical Hilbert space. It turns out that in the massless case
corresponding to Einstein's theory 
$
{\cal H}_{\rm phys}
$ 
is trivial. This reflects the
well-known fact mentioned above that Einstein's theory has no graviton states in 2+1 dimensions.
In the massive gauge theory 
$
{\cal H}_{\rm phys}
$ 
is non-trivial, for fixed momentum there exist three physical modes. This remains true in the limit 
$
m\to 0
$ 
so that we have a non-trivial theory in both cases. In section 3 we construct a concrete Hilbert space
representation in order to show that all operators are well defined. This representation is
chosen in such a way that the massless limit $m\to 0$ is smooth. In section 4 we
derive the coupling from the so-called descent equations which are a generalization of
causal gauge invariance (1.2). The even-parity sector can be treated in exactly the same way as in
$
3+1
$
dimensions. We carry through the descent procedure in the odd-parity case in all details. We recover the gravitational Chern-Simons coupling. Together with the even-parity coupling
we obtain the so-called ''cosmological topologically massive
gravity``[11]. From the point of view of quantum gauge theory the treatment of this theory 
in the literature is incomplete because the vector-graviton field is lacking. 
In the conclusions we point out further differences to the classical theory.

\newpage

\section{Massless and massive gravity in 2+1 dimensions}

As we have said in the Introduction, we use the framework from \cite{8}-\cite{16} which works fine for the four-dimensional case. We must first see if the same framework works in three dimensions. We use the
same convention as in 3+1 dimensions as far as possible. Then many results of the 4-dimensional theory
can be taken over without change. The basic free quantum field in massive gravity is a symmetric tensor field 
$
h^{\mu\nu}(x)
$
satisfying the Klein-Gordon equation
\be
(\Box+m^2)h^{\mu\nu}=0.
\ee
It is quantized according to
\be
[h^{\alpha\beta}(x), h^{\mu\nu}(y)]=-{i\over 2} (\eta^{\alpha\mu}\eta^{\beta\nu}+\eta^{\alpha\nu}
\eta^{\beta\mu}-\eta^{\alpha\beta}\eta^{\mu\nu})D_{m}(x-y),
\label{hh}
\ee
where 
$
\eta^{\mu\nu}
$ is the Minkowski tensor with diagonal elements
$
(1,-1,-1)
$.
$
D_m
$
is the 2+1-di\-men\-sional Jordan-Pauli distribution with mass $m$. To make 
$
h^{\mu\nu}
$ 
a gauge field we must introduce ghost and antighost fields with the same mass
\bea
(\Box+m^2)u^\mu=0=(\Box+m^2)\tilde u^\mu.
\nonumber
\eea
These fields are quantized with anti-commutators
\be
\{u^\mu(x),\tilde u^\nu(y)\}=i\eta^{\mu\nu}D_{m}(x-y)
\ee
and all other anti-commutators vanishing.

Then we can define the gauge variations
\be
d_Q h^{\mu\nu}=[Q,h^{\mu\nu}]=-{i\over 2} (\d^\nu u^\mu+\d^\mu u^\nu
-\eta^{\mu\nu}\d_\alpha u^\alpha)
\ee
\be
d_Q u^\mu=\{Q,u\}=0.
\ee
The gauge variation of 
$
\tilde u^\mu
$ is non-trivial. Since 
$
d_Q
$ 
is nilpotent,
$
d_Q^2=0
$,
we must introduce a vector field 
$
v^\mu(x)
$ 
with the same mass
\bea
(\Box+m^2)v^\mu=0
\nonumber
\eea
which we call vector-graviton field or $v$-field for short. It is quantized according to
\be
~[v^\mu(x),\, v^\nu(y)]= {i\over 2} \eta^{\mu\nu}D_{m}(x-y).
\ee
This field appears in the gauge variation of 
$
\tilde u^\mu
$
\be
d_Q\tilde u^\mu =\{Q,\tilde u^\mu\}=i(\d_\nu h^{\mu\nu}+m v^\mu).
\ee
Finally
\be
d_Q v^\mu=[Q,v^\mu]=-{i\over 2} m u^\mu
\label{dqv}
\ee
It is not hard to verify nilpotency 
$
d_Q^2=0
$. 
Using the commutation rules above one can show that $Q$ is expressed in $x$-space as follows
\be
Q= \int\limits_{x^0=t} d^3x\,\Bigl[\d_\nu h^{\mu\nu}(x)+m 
v^\mu(x)\Bigl]{\dl}_0 u_\mu(x).
\label{Q}
\ee
These relations remain true in the massless case 
$
m=0
$, 
but the $v$-field is then completely skipped in ordinary gravity theory.

We now describe the one-particle Hilbert space as in ref. \cite{11} and \cite{13}. First we study the massless case. The generic form of a state 
$
\Psi \in {\cal H}^{(1)} \subset {\cal H}
$
from the one-particle Hilbert subspace is
\be
\Psi = \left[ \int f_{\mu\nu}(x) h^{\mu\nu}(x) + \int g^{(1)}_{\mu}(x) u^{\mu}(x) 
+ \int g^{(2)}_{\mu}(x) \tilde{u}^{\mu}(x) \right] \Omega
\ee
with test functions
$
f_{\mu\nu}, g^{(1)}_{\mu}, g^{(2)}_{\mu}
$
verifying the wave equation; we can also suppose that
$
f_{\mu\nu}
$
is symmetric; we denote
$
f \equiv \eta^{\mu\nu}~f_{\mu\nu}
$.

The kernel of the gauge charge operator $Q$ (restricted to one-particle states) is given by states of the form
\be
\Psi = \left[ \int f_{\mu\nu}(x) h^{\mu\nu}(x) + \int g_{\mu}(x) u^{\mu}(x) \right] \Omega
\label{kerQ-0}
\ee
with $g_{\mu}$ arbitrary and 
$
f_{\mu\nu}
$
constrained by the condition 
$
\partial^{\nu}f_{\mu\nu} = {1\over 2}~\partial_{\mu}f;
$
so the elements of
$
{\cal H}^{(1)} \cap Ker(Q)
$
are in one-one correspondence with couples of test functions
$
[f_{\mu\nu}, g_{\rho}]
$
with the transversality condition on the first entry.

Now, a generic element
$
\Psi^{\prime} \in {\cal H}^{(1)} \cap Ran(Q)
$
has the form 
\be
\Psi^{\prime} = Q\Phi = \left[
- {1\over 2} \int (\partial_{\mu}g^{\prime}_{\nu} + \partial_{\nu}g^{\prime}_{\mu})(x) h^{\mu\nu}(x) 
+ \int \left(\partial^{\nu}g^{\prime}_{\mu\nu} 
- {1\over 2}~\partial_{\mu}g^{\prime}\right)(x) u(x) \right] \Omega
\label{ranQ-0}
\ee
with 
$
g^{\prime} = \eta^{\mu\nu}g^{\prime}_{\mu\nu}
$,
so if
$
\Psi \in {\cal H}^{(1)} \cap Ker(Q)
$
is indexed by the couple 
$
[f_{\mu\nu}, g_{\rho}]
$
then 
$
\Psi + \Psi^{\prime}
$
is indexed by the couple
$
\left[
f_{\mu\nu} - {1\over 2}~(\partial_{\mu}g^{\prime}_{\nu} + \partial_{\nu}g^{\prime}_{\mu}), 
g_{\mu} + \left( \partial^{\nu}g^{\prime}_{\mu\nu} 
- {1\over 2}~\partial_{\mu}g^{\prime}\right)\right].
$
If we take 
$
g^{\prime}_{\mu\nu}
$
conveniently we can make 
$
g_{\mu} = 0
$
and if we take 
$
g^{\prime}_{\mu}
$
convenient we can make 
$
f = 0;
$
in this case we have the transversality condition 
$
\partial^{\nu}f_{\mu\nu} = 0.
$
It follows that the equivalence classes from
$
({\cal H}^{(1)} \cap Ker(Q))/({\cal H}^{(1)} \cap Ran(Q))
$ 
are indexed by wave functions
$
f_{\mu\nu}
$
verifying the conditions of transversality and tracelessness
$
\partial^{\nu}f_{\mu\nu} = 0,~f = 0.
$

We go in the momentum space and choose a Lorentz frame such that 
$
P = (1,0,1).
$
Then the Fourier transform
$
\tilde{f}_{\mu\nu}(P)
$
is restricted by the two conditions above (transversality and tracelessness) and we have the non-null elements of the tensor given by two free parameters:
\bea
\tilde{f}_{00}(P) = \tilde{f}_{22}(P) = - \tilde{f}_{02}(P) = \alpha
\nonumber \\
\tilde{f}_{01}(P) = \tilde{f}_{12}(P) = \beta
\eea
with 
$
\alpha
$
and
$
\beta
$
two arbitrary complex numbers. Now if we compute the value of the ``scalar product''
\be
<\tilde f,\tilde f> = \tilde f_{\mu\nu}^{*}\tilde f^{\mu\nu}
\ee
for the previous values we get $0$. So there is no way to construct the one-particle Hilbert space as in the four-dimensional case.

The situation changes drastically in the massive case. The generic form of a state 
$
\Psi \in {\cal H}^{(1)} \subset {\cal H}
$
from the one-particle Hilbert subspace is
\be
\Psi = \left[ \int f_{\mu\nu}(x) h^{\mu\nu}(x) + \int g^{(1)}_{\mu}(x) u^{\mu}(x) 
+ \int g^{(2)}_{\mu}(x) \tilde{u}^{\mu}(x) + \int h_{\mu}(x) v^{\mu}(x) \right] \Omega
\ee
with test functions
$
f_{\mu\nu}, g^{(1)}_{\mu}, g^{(2)}_{\mu}, h_{\mu}
$
verifying the Klein-Gordon equation; we can also suppose that
$
f_{\mu\nu}
$
is symmetric. Now the elements
$
\Psi \in {\cal H}^{(1)} \cap Ker(Q)
$
from the kernel of the gauge charge operator are of the form
\be
\Psi = \left[ \int f_{\mu\nu}(x) h^{\mu\nu}(x) + \int g_{\mu}(x) u^{\mu}(x) 
+ {2\over m}~\int
\left(\partial^{\nu}f_{\mu\nu} - {1\over 2}~\partial_{\mu}f\right)(x) v^{\mu}(x)\right] \Omega
\label{kerQ-m}
\ee
with 
$g_{\mu}$ 
and 
$
f_{\mu\nu}
$
arbitrary so
$
\Psi \in {\cal H}^{(1)} \cap Ker(Q)
$
is indexed by couples of test functions
$
[f_{\mu\nu},g_{\mu}].
$
Now, a generic element
$
\Psi^{\prime} \in {\cal H}^{(1)} \cap Ran(Q)
$
has the form 
\be
\Psi^{\prime} = Q\Phi = \left[
- {1\over 2} \int (\partial_{\mu}g^{\prime}_{\nu} + \partial_{\nu}g^{\prime}_{\mu})(x) h^{\mu\nu}(x) 
+ \int \left(\partial^{\nu}g^{\prime}_{\mu\nu} 
- {1\over 2}~\partial_{\mu}g^{\prime} - {m\over 2} h^{\prime}_{\mu}\right)(x) u^{\mu}(x) \right] \Omega
\label{ranQ-m}
\ee
with 
$
g^{\prime} = \eta^{\mu\nu}g^{\prime}_{\mu\nu}
$
so if
$
\Psi \in {\cal H}^{(1)} \cap Ker(Q)
$
is indexed by the couple 
$
[f_{\mu\nu}, g_{\rho}]
$
then 
$
\Psi + \Psi^{\prime}
$
is indexed by the couple
$
\left[
f_{\mu\nu} - {1\over 2}~(\partial_{\mu}g^{\prime}_{\nu} + \partial_{\nu}g^{\prime}_{\mu}), 
g_{\mu} + \left( \partial^{\nu}g^{\prime}_{\mu\nu} 
- {1\over 2}~\partial_{\mu}g^{\prime} - {m\over 2} h^{\prime}_{\mu}\right)\right].
$
If we take 
$
h^{\prime}_{\mu}
$
conveniently we can make 
$
g_{\mu} = 0
$
and if we take 
$
g^{\prime}_{\mu}
$
convenient we can make 
\be
\partial^{\nu}f_{\mu\nu} - {1\over 2}~\partial_{\mu}f = 0.
\ee

As above we consider a Lorentz reference frame where
$
P = (m,0,0)
$
and we get from the condition above that the non-null elements of the expression
$
\tilde{f}_{\mu\nu}(P)
$
depend on three free parameters:
\bea
\tilde{f}_{11}(P) = \alpha, \quad \tilde{f}_{22}(P) = \beta, \qquad  \tilde{f}_{12}(P) = \gamma, \qquad
\tilde{f}_{00}(P) = \alpha + \beta
\eea

If we compute the value of the ``scalar product''
\be
<\tilde{f},\tilde{f}> = \tilde{f}_{\mu\nu}^{*} \tilde{f}^{\mu\nu}
\ee
for the previous values we get in this case
\be
<f,f> = |\alpha+\beta|^{2} + |\alpha|^{2} + |\beta|^{2} + 2 |\gamma|^{2}
\ee
which is positively defined and induces a well-defined scalar product on the physical Hilbert space
$
{\cal H}_{\rm phys}={\rm Ker}Q/{\rm Ran}Q.
$

Moreover, if we apply a rotation of angle
$
\phi
$
(which is an element of the stability group of the momentum $P$) to the expression 
$
\tilde{f}_{\mu\nu}(P)
$
we immediately obtain that the expression 
$
\alpha + \beta
$ 
is invariant (so it describes a spin $0$ particle) and the expressions
$
\alpha - \beta \pm 2 i \gamma
$
are transformed by a phase factor
$
e^{\pm 2 i \phi}
$
(so they describe two particles of spin $\pm 2$ respectively).

It follows that we have a good description for the massive spin $2$ particles in three dimension which is similar to the four-dimensional case. Another construction of the physical Hilbert space is given in the next Section. 
\newpage

\section{Representation in momentum space}

To understand the gauge structure better we construct a Hilbert space representation
of the massive 2+1-dimensional theory. For this purpose we express the
various fields by means of emission and absorption operators. In doing so we have
to introduce a positive definite scalar product which breaks Lorentz invariance
but defines the topology of the
big Fock space of physical and unphysical states and the adjoint operators.
It is well known that this Hilbert structure is not unique \cite{19}, we shall chose it
in such a way that we get a smooth massless limit 
$
m\to 0
$. 
We follow
the discussion of the 4-dimensional case as close as possible [10]. We decompose 
$
h^{\alpha\beta}
$ 
into its traceless part and the trace $h$
\be
h^{\alpha\beta}(x)=H^{\alpha\beta}(x)+{1\over 3}\eta^{\alpha\beta}
h(x).
\ee
>From (\ref{hh}) we obtain the following commutation relations
\be
[h(x),h(y)]={3i\over 2}D_{m}(x-y)
\label{hh1}
\ee
\be
[H^{\alpha\beta}(x), H^{\mu\nu}(y)]=-{i\over 2}\left(\eta^{\alpha\mu}\eta^{\beta\nu}+
\eta^{\alpha\nu}\eta^{\beta\mu}-{2\over 3}\eta^{\alpha\beta}\eta^{\mu\nu}\right)~D_{m}(x-y),
\label{hh2}
\ee
and
\be
[H^{\alpha\beta}(x), h(y)]=0.
\ee
It is easy to verify that the fields in (\ref{hh1}) (\ref{hh2}) can be represented as follows
\be
H^{\alpha\beta}(x)= (2\pi)^{-1}\int{d^2k\over\sqrt{2E_k}}\,\Bigl( a_{\alpha\beta}(\vec k)e^{-ikx}
+\eta^{\alpha\alpha}\eta^{\beta\beta}a_{\alpha\beta}^+(\vec k)e^{ikx}\Bigl).
\ee
Here 
$
E_k=\sqrt{\vec k^2+m^2}
$,
$
a_{\alpha\beta}=a_{\beta\alpha}
$ 
is symmetric and satisfies the commutation relation
\be
[a_{\alpha\beta}(\vec k), a_{\mu\nu}^+(\vec k')]=\eta^{\alpha\alpha}
\eta^{\beta\beta} \left(\eta^{\alpha\mu}\eta^{\beta\nu}+
\eta^{\alpha\nu}\eta^{\beta\mu}-{2\over 3}\eta^{\alpha\beta}\eta^{\mu\nu}\right)\delta(\vec k-\vec k').
\ee
The trace part is given by
\be
h(x)=(2\pi)^{-1}\int{d^2k\over\sqrt{2E_k}}\Bigl(a(\vec k)e^{-ikx}-a^+(\vec k)e^{ikx}\Bigl)
\ee
with
\be
[a(\vec k), a^+(\vec k')]={3\over 2}\delta(\vec k-\vec k').
\ee
Since the right-hand side is positive, the $h$-sector of Fock space can be
constructed in the usual way by applying products of $a^+$'s to the
vacuum.

The situation is not so simple in the $H$-sector because the righthand side of (3.6) is not a diagonal matrix. We perform a linear transformation of the diagonal operators 
$
a_{\alpha\alpha}
$ 
and 
$
a_{\alpha\alpha}^+
$
in such a way that the new operators are usual annihilation and creation operators satisfying
\be
[\tilde a_{\alpha\alpha}(\vec k),\tilde a_{\beta\beta}^+(\vec k')]
=\delta_{\alpha\beta}\delta(\vec k-\vec k').
\ee
This is achieved by the following transformation:
\bea
a_{00}=\sqrt{2\over 3}(\tilde a_{11}+\tilde a_{22}+\tilde a_{33})
\nonumber \\
a_{11}=\alpha_1\tilde a_{11}+\alpha_2\tilde a_{22}
\nonumber \\
a_{22}=\alpha_2\tilde a_{11}+\alpha_2\tilde a_{22},
\label{a}
\eea
with
\be
\alpha_1={1\over\sqrt{6}}-{1\over\sqrt{2}},\quad
\alpha_2={1\over\sqrt{6}}+{1\over\sqrt{2}}.
\ee
We note that 
$
\tilde a_{00}
$ 
does not appear because one pair of absorption and emission operators is superfluous due to the trace condition
$
H^\alpha\,_\alpha=0
$. 
In fact, from (3.10) we see
\bea
\sum_{j=1}^2a_{jj}=a_{00}.
\nonumber
\eea
The Fock representation can now be constructed as usual by means of
$
\tilde a_{11}^+,\tilde a_{22}^+
$ 
and 
$
a_{\alpha\beta}^+
$ 
with 
$
\alpha\ne\beta
$.

The other fields have the following representation in terms of emission and absorption operators:
\bea
u^\mu(x)=(2\pi)^{-1}\int{d^2k\over\sqrt{2E_k}}\,\Bigl(c_2^\mu(\vec
k)e^{-ikx}-\eta^{\mu\mu}c_1^\mu(\vec k)^+e^{ikx}\Bigl)
\nonumber \\
\tilde u^\mu(x)=(2\pi)^{-1}\int{d^2k\over\sqrt{2E_k}}\,
\Bigl(-c_1^\mu(\vec k)e^{-ikx}-\eta^{\mu\mu}c_2^\mu
(\vec k)^+e^{ikx}\Bigl)
\eea
\be
v^\mu(x)=(2\pi)^{-1}\int {d^2k\over 2\sqrt{E_k}}\,\Bigl(b^\mu(\vec
k)e^{-ikx}-\eta^{\mu\mu}b^\mu(\vec k)^+e^{ikx}\Bigl)
\ee
with the following (anti)commutation relations
\be
\{c_j^\mu(\vec k),c_l^\nu(\vec k')^+\}=\delta_{jl}\delta^\mu_{\nu} 
\delta^3(\vec k-\vec k'),
\ee
\be
[b^\mu(\vec k), b^\nu(\vec k')^+]=\delta^\mu_\nu\delta^3(\vec k-\vec k').
\ee
Then the gauge charge $Q$ (\ref{Q}) can be written in momentum space as follows
\be 
Q=\int d^3k\,\Bigl(A^\alpha(\vec k)^+c_2^\gamma(\vec k)-B^\alpha(\vec k)
c_1^\gamma(\vec k)^+\Bigl)\eta_{\alpha\gamma},
\ee
where
\be
A^\alpha=\eta^{\alpha\alpha}\eta^{\beta\beta}a^{\alpha\beta}(\vec k)k^\beta-
{k^\alpha\over 4}d(\vec k)-im_1\eta^{\alpha\alpha}b^\alpha
\ee
\be
B^\alpha=(a^{\alpha\beta}(\vec k)k_\beta+
{k^\alpha\over 4}d(\vec k)+im_1b^\alpha)\eta^{\alpha\alpha},
\ee
\be
m_1={m\over\sqrt{2}}.
\ee
The adjoint is given by
\be
Q^+=\int d^2k\,\Bigl(c_2^\beta(\vec k)^+A^\alpha(\vec k)-
c_1^\beta(\vec k)B^\alpha(\vec k)^+\Bigl)\eta_{\alpha\beta}.
\ee

The physical Hilbert space can be expressed by means of the gauge charge $Q$ in the following equivalent form
\be
{\cal H}_{\rm phys}={\rm Ker}(QQ^++Q^+Q).
\label{kernel}
\ee
We must study the selfadjoint operator
\bea
\{Q,Q^+\}=\int d^3k\,d^3k'\,\Bigl(A^\alpha(\vec k)^+A^\beta(\vec k')
\{c_2^\gamma(\vec k),c_2^\delta(\vec k')^+\}
\nonumber \\
+B^\beta(\vec k')^+B^\alpha(\vec k)\{c_1^\delta(\vec k'),c_1^\gamma
(\vec k)^+\}+c_2^\delta(\vec k')^+c_2^\gamma(\vec k)[A^\beta(\vec k'), 
A^\alpha(\vec k)^+]
\nonumber \\ 
+c_1^\gamma(\vec k)^+c_1^\delta(\vec k')[B^\alpha(\vec k),
B^\beta(\vec k')^+]\Bigl)\eta_{\alpha\gamma}\eta_{\beta\delta}.
\eea
We restrict to the graviton sector because the ghost sector is totally unphysical:
\be
\{ Q,Q^+\}\vert_{\rm graviton}
=\int d^3k\,\sum_{\alpha=0}^3\Bigl(A^{\alpha +}A^\alpha+
B^{\alpha +}B^\alpha\Bigl).
\label{QQ}
\ee
It is convenient to introduce time-like and space-like components:
\bea
A^0=k_0(a^{00}-a^0_\parallel-{a\over 3}-{im_1\over k_0}b^0),
\nonumber \\
A^j=k_0(-a^{0j}+a^j_\parallel-{k^j\over k_0}{a\over 3}+{im_1\over k_0}b^j),
\nonumber \\
B^0=k_0(a^{00}+a^0_\parallel+{a\over 3}+{im_1\over k_0}b^0),
\nonumber \\
B^j=k_0(-a^{0j}-a^j_\parallel-{k^j\over k_0}{a\over 3}-{im_1\over k_0}b^j),
\eea
where
\be
a_\parallel^\mu={k_j\over k_0}a^{\mu j}.
\ee

We choose a Lorentz frame where 
$
k^\mu=(k_0,0,k_2)
$ 
and substitute the diagonal operators
$
a^{\mu\mu}
$ 
by 
$
\tilde a_{jj}
$ 
(\ref{a}). Then we get for the integrand in (\ref{QQ})
\bea
\sum_{\alpha =0}^3\Bigl(A^{\alpha +}A^\alpha+B^{\alpha +}B^\alpha\Bigl)= 
2k_0^2\Bigl\{{2\over 3}(\tilde a_{11+}+\tilde a_{22+})(\tilde a_{11}+\tilde a_{22}) +{k_2^2\over k_0^2}a^{02+}a^{02}+
\nonumber \\ 
+{im_1k_2\over k_0^2}a^{02+}b_0+{1\over 3}{k_2\over k_0}a^{02+}a+{1\over 9}a^+a+
{im_1\over 3k_0}a^+b_0-{imk_2\over k_0^2}b_0^+a^{02}-{im\over 3k_0}b_0^+a+
\nonumber \\
+{m_1^2\over k_0^2}b_0^+b_0+a^{01+}a^{01}+a^{02+}a^{02}+{k_2\over 3k_0}a^{02+}a+
{k_2^2\over k_0^2}a^{12+}a^{12}+
\nonumber \\ 
+{k_2^2\over k_0^2}(\alpha_2\tilde a_{11}^++\alpha_1\tilde a_{22}^+)(\alpha_2\tilde a_{11}+ \alpha_1\tilde a_{22})+{im_1k_2\over k_0^2}a^{12+}b^1+
\nonumber \\ 
+{im_1k_2\over k_0^2}(\alpha_2\tilde a_{11}^++\alpha_1\tilde a_{22}^+)b^2
+{k_2\over 3k_0}a^+a^{02}+{k_2^2\over 9k_0^2}a^+a-{im_1k_2\over k_0^2}b^{1+}a^{12}-
\nonumber \\
-{im_1k_2\over k_0^2}b^{2+}(\alpha_2\tilde a_{11}^++\alpha_1\tilde a_{22}^+)+
{m_1^2\over k_0^2}(b^{1+}b^ 1+b^{2+}b^2)\Bigl\}.
\label{AA}
\eea
Since 
$
a^{11+}
$ 
does not appear inhere, the states 
$
a^{11+}\Omega
$ 
where
$
\Omega
$ 
is the Fock vacuum certainly belong to the kernel of (\ref{QQ}) and, hence, are in the physical subspace.

The quadratic form (\ref{AA}) can be represented in matrix notation
$
A^+XA
$ 
where 
$
A^+
$ 
stands for the emission operators
\be
A^+=(\tilde a_{11}^+,\tilde a_{22}^+,b_2^+,b_1^+,a_{12}^+,a_{02}^+,a^+,  b_0^+,a_{01}^+).
\ee
The matrix $X$ has block diagonal form with the following three submatrices:
\bea
X_0=\pmatrix{1+{k_2^2\over k_0^2}&{k_2\over 3k_0}&{im_1k_2\over k_0^2}\cr
{k_2\over 3k_0}&{1\over 9}+{k_2^2\over 9k_0^2}&{im_1\over 3k_0}\cr
-{imk_2\over k_0^2}&0&{m_1^2\over k_0^2}\cr},\quad
X_1=\pmatrix{{m^2\over k_0^2}&-{im_1k_2\over k_0^2}\cr
{im_1k_2\over k_0^2}&{k_2^2\over k_0^2}\cr}
\nonumber \\
X_2=\pmatrix{{2\over 3}+{\alpha_2^2k_2^2\over k_0^2}& {2\over 3}+{\alpha_1\alpha_2k_2^2 
\over k_0^2}&{im_1\alpha_2k_2}\over k_0^2\cr
{2\over 3}+{\alpha_1\alpha_2k_2^2\over k_0^2}&{2\over 3}+{\alpha_1^2k_2^2\over k_0^2}&
{im_1\alpha_1k_2\over k_0^2}\cr
-{im_1\alpha_2k_2\over k_0^2}&0&{m_1^2\over k_0^2}\cr}.
\eea
The kernel (\ref{kernel}) now consists of the null-vectors of these matrices. Only 
$
X_1
$ 
and 
$
X_2
$
have eigenvalue 0, the corresponding eigenvectors are
\bea
\psi_1=\Bigl(b_1^+-{im_1\over k_2}a^{12+}\Bigl)\Omega
\nonumber \\
\psi_2=\Bigl(b_2^++{im_1\over \sqrt{2}k_2}(\tilde a_2^+-\tilde a_1^+)\Bigl)\Omega.
\eea
In the limit 
$
m\to 0
$ 
these two physical states go over into the free vector-graviton states. Consequently, the physical modes of the massless theory are one transversal graviton state
$
a^{11+}\Omega
$ 
plus these two vector-graviton states. In the massive case there is some admixture of other graviton states (3.29).
\newpage
\section{Interaction from descent equations}

There are a number of ideas which must be used to determine in an unique way the expression of the interaction Lagrangian respecting the gauge invariance condition (\ref{gauge-inv}). First, because we are in three dimensions, power counting allows us to consider Wick polynomials of canonical dimension 
$
\omega(T), \omega(T^{\alpha}) \leq 6
$ 
and tri-linear in the fields and their derivatives. 

Next, we can obtain from (\ref{gauge-inv}) by a standard procedure the chain of relations
\bea
d_{Q}T^{\alpha} = i~\d_{\beta}T^{[\alpha\beta]}.
\nonumber \\
d_{Q}T^{[\alpha\beta]} = i~\d_{\gamma}T^{[\alpha\beta\gamma]}
\nonumber \\
d_{Q}T^{[\alpha\beta\gamma]} = 0
\label{descent-T}
\eea
where the carets emphasize complete antisymmetry. These relations can be solved starting from the last one (in top ghost number equal to $3$).  Going backwards in this chain of relations we are always reduced to solve co-cycle conditions of the type
$
d_{Q}C = 0.
$
This is the descent procedure.

The cohomology of the operator
$
d_{Q}
$
has been investigated in \cite{13} in the four-dimensional case. Because we have preserved the algebraic structure of the gauge charge operator the analysis from this reference remains unchanged: the space dimension plays no r\^ole. One must determine the invariants with respect to the gauge charge i.e. solutions of the equation
$
d_{Q}C = 0
$
which cannot be expressed as co-boundaries i.e. in the form
$
C = d_{Q}B
$
and of canonical dimension bounded by some integer $n$. We denote by 
$
{\cal P}^{n}, Z_{Q}^{n}, B_{Q}^{n}
$ 
the space of cochains, cocycles and coboundaries respectively and we require
$
\omega(B_{Q}^{n}) \leq n - 1
$.

In the massless case these invariants are
$
u_{\mu}
$,
the antisymmetric first-order derivative
\be
u_{[\mu\nu]} \equiv {1\over 2}~(\d_{\mu}u_{\nu} - \d_{\nu}u_{\mu})
\ee
and the (linear) Riemann tensor and its derivatives. One defines it as follows: first we introduce the {\it Christoffel symbols} according to:
\be
\Gamma_{\mu;\nu\rho} \equiv \d_{\rho}\hat{h}_{\mu\nu} + \d_{\nu}\hat{h}_{\mu\rho} - \d_{\mu}\hat{h}_{\nu\rho}
\ee 
where
\be
h \equiv \eta^{\mu\nu}h_{\mu\nu} \qquad
\hat{h}_{\mu\nu} \equiv h_{\mu\nu} - \eta_{\mu\nu}~h
\ee
and then the Riemann tensor is:
\be
R_{\mu\nu;\rho\sigma} \equiv \d_{\rho}\Gamma_{\mu;\nu\sigma} - (\rho \leftrightarrow \sigma).
\ee
One must eliminate in the systematic way all traces from the derivatives of the Riemann tensor to obtain true invariants; the traces are coboundaries. Then one can prove that any cocycle is cohomologous to a Wick polynomial in the invariants.

In the massive case some invariants are lost:
$
u_{\mu}
$ 
and
$
u_{[\mu\nu]}
$
become coboundaries by (\ref{dqv}) and a new invariant appears:
\bea
\phi_{\mu\nu} \equiv 
- \d_{\mu}v_{\nu} - \d_{\nu}v_{\mu} + \eta_{\mu\nu} \d_{\rho}v^{\rho} + m~h_{\mu\nu}
\nonumber \\
\phi \equiv \eta^{\mu\nu}~\phi_{\mu\nu}
\eea
and their derivatives. (Again one must conveniently eliminate the traces of the various derivative). However, the situation is still more involved in the massive case. Any cocycle is cohomologous to an expression of the form
\be
p_{1} + d_{Q}p_{2}
\label{co}
\ee 
where 
$
p_{1}
$
depends only on the invariants and 
$
p_{2}
$ 
is a Wick polynomial of canonical dimension equal to $n$ and such that
$
d_{Q}p_{2}
$
has also canonical dimension $n$. This is because the expression
$
d_{Q}p_{2}
$
is a cocycle in any canonical dimension bigger that $n$ i.e. in
$
{\cal P}^{m},~m > n
$
but is not a co-boundary in 
$
{\cal P}^{n}.
$

A final idea is related to the fact that we are working in 3 dimensions so we have the Lorentz invariant and completely antisymmetric tensor
$
\epsilon_{\mu\nu\rho}
$
which allows us to trade a couple of antisymmetric indices for only one index. For instance, instead of
$
u_{[\mu\nu]}
$
we prefer to work with the variables
\be
\lambda_{\mu} \equiv {1\over 2}~\epsilon_{\mu\rho\sigma}~u^{[\rho\sigma]} \quad \Leftrightarrow \quad
u_{[\mu\nu]} = \epsilon_{\mu\nu\rho}~\lambda^{\rho}.
\ee
We also define the expressions
\be
\Gamma_{\mu;\nu} \equiv {\epsilon_{\mu}^{\cdot}}^{\rho\sigma}~\Gamma_{\rho;\sigma\nu}
\ee
and observe that
\be
\d_{\nu}\lambda_{\mu} = - {i\over 2}~d_{Q} \Gamma_{\mu;\nu}.
\label{lambda-gamma}
\ee
Finally we define
\be
V_{\mu} \equiv \epsilon_{\mu\rho\sigma}~\d^{\rho}v^{\sigma} \quad \Leftrightarrow \quad
\d_{\mu}v_{\nu} - \d_{\nu}v_{\mu} = \epsilon_{\mu\nu\rho}~V^{\rho}
\ee
so that we have
\be
d_{Q}V_{\mu} = i m \lambda_{\mu}.
\ee

We are ready to start the descent procedure. We must start with the expression
$
T^{[\alpha\beta\gamma]}
$
of the descent system (\ref{descent-T}). One must use the limitations on the canonical dimension (
$
\omega(T^{\alpha\beta\gamma}) \leq 6
$),
ghost number (
$
gh(T^{\alpha\beta\gamma}) = 3
$)
and complete antisymmetry. In canonical dimension $5$ we have the same expression as in  four dimensions. The descent procedure goes though in exactly the same way as in \cite{13} and \cite{15} so we obtain in the end the same interaction terms related to the Einstein-Hilbert Lagrangian (see the Conclusions). However, in canonical dimension $6$ a new expression appears:
\be
T^{[\alpha\beta\gamma]} = c_{0}~\epsilon^{\alpha\beta\gamma}~\epsilon^{\mu\nu\rho}~\lambda_{\mu}~\lambda_{\nu}~\lambda_{\rho} 
+ i~d_{Q}B^{[\alpha\beta\gamma]}
\label{t3}
\ee
The first term can be rewritten (up to a constant) as
$
d_{Q}(\epsilon^{\alpha\beta\gamma}~\epsilon^{\mu\nu\rho}~V_{\mu}~\lambda_{\nu}~\lambda_{\rho})
$;
the expression in the bracket is of canonical dimension $6$ so it is of the form
$
d_{Q}p_{2}
$
from (\ref{co}). For simplicity we take
$
c_{0} = 1
$
because anyway we can rescale the final solution by a constant. We substitute this in the second equation (\ref{descent-T}), use (\ref{lambda-gamma}) and obtain
\bea
d_{Q}\left(T^{[\alpha\beta]} - {3\over 2}~\epsilon^{\alpha\beta\gamma}~\lambda_{\mu}~\lambda_{\nu}~\Gamma_{\rho;\gamma} 
+ \d_{\gamma}B^{[\alpha\beta\gamma]} \right) = 0.
\eea

Using the description of the cocycles for
$
d_{Q}
$
we get from here:
\be
T^{[\alpha\beta]} = {3\over 2}~\epsilon^{\alpha\beta\gamma}~\epsilon^{\mu\nu\rho}~\lambda_{\mu}~\lambda_{\nu}~\Gamma_{\rho;\gamma} 
+ i d_{Q}B^{[\alpha\beta]} - \d_{\gamma}B^{[\alpha\beta\gamma]} + T^{[\alpha\beta]}_{0}
\label{t2}
\ee
where the last term
$
T^{[\alpha\beta]}_{0}
$
depends only on the invariants. Now we substitute the preceding expression in the first equation (\ref{descent-T}) and after some computations we arrive at
\bea
d_{Q} \Bigl[ T^{\alpha} + {3\over 4}~\epsilon^{\alpha\beta\gamma}~\epsilon^{\mu\nu\rho}~\lambda_{\mu}~\Gamma_{\nu;\beta}~\Gamma_{\rho;\gamma}
\nonumber \\
+ 3~\epsilon^{\mu\nu\rho}~\lambda_{\mu}~\lambda_{\nu}~(\d_{\rho}\tilde{u}^{\alpha} + \d^{\alpha}\tilde{u}_{\rho} 
- \delta^{\alpha}_{\rho}~\d_{\sigma}\tilde{u}^{\sigma})
+ 3~\epsilon_{\mu\nu\rho}~\lambda^{\mu}~V^{\nu}~\phi^{\rho\alpha} + \d_{\beta}B^{[\alpha\beta]} \Bigl] 
\nonumber \\
= \d_{\beta}~T^{[\alpha\beta]}_{0} 
\label{x} 
\eea
It is easy to verify this by computing 
$
d_{Q}
$ 
of the terms in the bracket and comparing with 
$
\d_{\beta}T^{\alpha\beta}
$
from (\ref{t2}).

It follows that the divergence in the right hand side must be a coboundary. It is rather straightforward to write a general ansatz for the expression
$
T^{[\alpha\beta]}_{0}
$ 
and use this condition. The result is that
$
T^{[\alpha\beta]}_{0}
$
is a relative cocycle i.e. an expression of the form
\be
T^{[\alpha\beta]}_{0} = i d_{Q}B^{[\alpha\beta]}_{0} - \d_{\gamma}B^{[\alpha\beta\gamma]}_{0} 
\ee
so we can get rid of the last term in (\ref{t2}) if we redefine
$
B^{[\alpha\beta]}
$
and
$
B^{[\alpha\beta\gamma]}
$
properly. We are left with
\be
T^{[\alpha\beta]} = {3\over 2}~\epsilon^{\alpha\beta\gamma}~\epsilon^{\mu\nu\rho}~\lambda_{\mu}~\lambda_{\nu}~\Gamma_{\rho;\gamma} 
+ i d_{Q}B^{[\alpha\beta]} - \d_{\gamma}B^{[\alpha\beta\gamma]}
\label{t2a}
\ee
and (\ref{x}) becomes
\bea
d_{Q} \Bigl[ T^{\alpha} + {3\over 4}~\epsilon^{\alpha\beta\gamma}~\epsilon^{\mu\nu\rho}~\lambda_{\mu}~\Gamma_{\nu;\beta}~\Gamma_{\rho;\gamma}
\nonumber \\
+ 3~\epsilon^{\mu\nu\rho}~\lambda_{\mu}~\lambda_{\nu}~(\d_{\rho}\tilde{u}^{\alpha} + \d^{\alpha}\tilde{u}_{\rho} 
- \delta^{\alpha}_{\rho}~\d_{\sigma}\tilde{u}^{\sigma})
+ 3~\epsilon_{\mu\nu\rho}~\lambda^{\mu}~V^{\nu}~\phi^{\rho\alpha} + \d_{\beta}B^{[\alpha\beta]} \Bigl] = 0
\eea   

Using again the cohomology of the operator
$
d_{Q}
$
we obtain that
\bea
T^{\alpha} = - {3\over 4}~\epsilon^{\alpha\beta\gamma}~\epsilon^{\mu\nu\rho}~\lambda_{\mu}~\Gamma_{\nu;\beta}~\Gamma_{\rho;\gamma}
\nonumber \\
- 3~\epsilon^{\mu\nu\rho}~\lambda_{\mu}~\lambda_{\nu}~(\d_{\rho}\tilde{u}^{\alpha} + \d^{\alpha}\tilde{u}_{\rho} 
- \delta^{\alpha}_{\rho}~\d_{\sigma}\tilde{u}^{\sigma})
- 3~\epsilon_{\mu\nu\rho}~\lambda^{\mu}~V^{\nu}~\phi^{\rho\alpha} + T^{\alpha}_{0}
\nonumber \\
+ i~d_{Q}B^{\alpha} - \d_{\beta}B^{[\alpha\beta]}
\label{t1}
\eea
where
$
T^{\alpha}_{0}
$
depends only on the invariants. A generic expression is
\bea
T^{\alpha}_{0} = a_{1}~\lambda^{\alpha}~\phi^{2} + a_{2}~\lambda^{\alpha}~\phi^{\rho\sigma}~\phi_{\rho\sigma}
+ a_{3}~\phi^{\alpha\beta}~\phi_{\beta\mu}~\lambda^{\mu} + a_{4}~\phi^{\alpha\beta}~\phi~\lambda_{\beta}.
\eea
We now substitute the expression (\ref{t1}) in the basic equation (\ref{gauge-inv}). For this purpose we calculate
$
\d_{\alpha}T^{\alpha}
$
and write it as a coboundary
$
d_{Q}(\cdots)
$
plus a rest. Then we obtain
\bea
d_{Q} \Bigl[ T + {1\over 8}~\epsilon^{\alpha\beta\gamma}~\epsilon^{\mu\nu\rho}~\Gamma_{\mu;\alpha}~\Gamma_{\nu;\beta}~\Gamma_{\rho;\gamma}
- 3~\epsilon^{\mu\nu\rho}~\lambda_{\mu}~\Gamma_{\nu;\alpha}~(\d_{\rho}\tilde{u}^{\alpha} + \d^{\alpha}\tilde{u}_{\rho}
- \delta^{\alpha}_{\rho}~\d_{\sigma}\tilde{u}^{\sigma})
\nonumber \\
+ {3\over 2}~\epsilon_{\mu\nu\rho}~\Gamma^{\mu;\alpha}~V^{\nu}~{\phi_{\alpha}^{\cdot}}^{\rho}
- 6~m~\epsilon_{\mu\nu\rho}~\lambda^{\mu}~V^{\nu}~\tilde{u}^{\rho} - t + \d_{\alpha}B^{\alpha} \Bigl]
\nonumber \\
= \left(a_{3} + {3\over 2}\right) \lambda_{\mu} \d^{\rho}\phi^{\mu\sigma}~\phi_{\rho\sigma} 
+ \left(2 a_{2} - {3\over 2}\right)~\lambda_{\mu}~\d^{\mu}\phi^{\rho\sigma}~\phi_{\rho\sigma}
\nonumber \\
+ \left(a_{4} - {3\over 2}\right) \lambda_{\mu} \d_{\mu}\phi~\phi^{\mu\nu} 
+ \left(2 a_{1} + {3\over 2}\right)~\lambda_{\mu}~\d^{\mu}\phi~\phi
\eea
where
\be
t \equiv a_{3}~\left( m \tilde{u}^{\beta} \phi_{\beta\mu}~\lambda^{\mu} 
+ {1 \over 2}~\phi^{\alpha\beta}~{\phi_{\beta}}^{\cdot\mu}~\Gamma_{\mu;\alpha}\right)
+ a_{4}~\left(m \tilde{u}^{\beta} \phi~\lambda^{\beta} + {1 \over 2}~\phi^{\alpha\beta}~\phi~\Gamma_{\beta;\alpha}\right)
\ee
It follows that we must choose
\be
a_{1} = - {3 \over 4},~~a_{2} = {3 \over 4},~~a_{3} = - {3 \over 2},~~a_{4} = {3 \over 2}.
\ee
and the preceding relation is
\bea
d_{Q} \Bigl[ T + {1\over 8}~\epsilon^{\alpha\beta\gamma}~\epsilon^{\mu\nu\rho}~\Gamma_{\mu;\alpha}~\Gamma_{\nu;\beta}~\Gamma_{\rho;\gamma}
- 3~\epsilon^{\mu\nu\rho}~\lambda_{\mu}~\Gamma_{\nu;\alpha}~(\d_{\rho}\tilde{u}^{\alpha} + \d^{\alpha}\tilde{u}_{\rho} 
- \delta^{\alpha}_{\rho}~\d_{\sigma}\tilde{u}^{\sigma})
\nonumber \\
+ {3\over 2}~\epsilon_{\mu\nu\rho}~\Gamma^{\mu;\alpha}~V^{\nu}~{\phi_{\alpha}^{\cdot}}^{\rho}
- 6~m~\epsilon_{\mu\nu\rho}~\lambda^{\mu}~V^{\nu}~\tilde{u}^{\rho} - t + \d_{\alpha}B^{\alpha} \Bigl] = 0
\eea
where now
\be
t \equiv ~{3 \over 2}~\Bigl( - m \tilde{u}^{\beta} \phi_{\beta\mu}~\lambda^{\mu} 
- {1 \over 2}~\phi^{\alpha\beta}~{\phi_{\beta}}^{\cdot\mu}~\Gamma_{\mu;\alpha}
+ m~\tilde{u}^{\beta} \phi~\lambda^{\beta} + {1 \over 2}~\phi^{\alpha\beta}~\phi~\Gamma_{\beta;\alpha}\Bigl)
\label{tt}
\ee

Finally, we apply once more the description of the cohomology of the operator
$
d_{Q}
$
and get
\bea
T = - {1\over 8}~\epsilon^{\alpha\beta\gamma}~\epsilon^{\mu\nu\rho}~\Gamma_{\mu;\alpha}~\Gamma_{\nu;\beta}~\Gamma_{\rho;\gamma}
\nonumber \\
+ 3~\epsilon^{\mu\nu\rho}~\lambda_{\mu}~\Gamma_{\nu;\alpha}~(\d_{\rho}\tilde{u}^{\alpha} + \d^{\alpha}\tilde{u}_{\rho} 
- \delta^{\alpha}_{\rho}~\d_{\sigma}\tilde{u}^{\sigma})
\nonumber \\
- {3\over 2}~\epsilon_{\mu\nu\rho}~\Gamma^{\mu;\alpha}~V^{\nu}~{\phi_{\alpha}^{\cdot}}^{\rho}
+ 6~m~\epsilon_{\mu\nu\rho}~\lambda^{\mu}~V^{\nu}~\tilde{u}^{\rho} + t + T_{0}
\nonumber \\
+ i~d_{Q}B - \d_{\alpha}B^{\alpha}
\label{t}
\eea
where the expression 
$
T_{0}
$
depends only on invariants. The generic form is
\be
T_{0} = c_{1}~\phi^{\mu\nu}~\phi_{\nu\rho}~{\phi_{\mu}^{\cdot}}^{\rho} + c_{2}~\phi^{\mu\nu}~\phi_{\mu\nu}~\phi + c_{3}~\phi^{3}.
\label{t0}
\ee

The final result is given by the formula (\ref{t}) together with (\ref{tt}) and (\ref{t0}). One can prove that 
\be
\epsilon^{\alpha\beta\gamma}~\epsilon^{\mu\nu\rho}~\Gamma_{\mu;\alpha}~\Gamma_{\nu;\beta}~\Gamma_{\rho;\gamma}
= 8~\epsilon^{\alpha\beta\gamma}~\Gamma^{\nu}_{\rho\alpha}~\Gamma^{\rho}_{\mu\beta}~\Gamma^{\mu}_{\nu\gamma}
\ee 
which is the well-known tri-linear part of the Chern-Simons coupling for gravity in three dimensions. The second line in (\ref{t}) is a ghost - antighost - graviton coupling, and the third line gives the coupling to the vector-graviton field. Without these couplings quantum gauge invariance is violated.
\newpage
\section{Conclusions}

We have derived the coupling of 2+1-dimensional massive quantum gauge theory on Minkowski
space from descent equations without using any classical Lagrangian. The complete coupling $T$
including the parity conserving part is equal to
\bea
T=\kappa\Bigl\{-h^{\alpha\beta}\d_\alpha h\d_\beta h+2h^{\alpha\beta}\d_\alpha h_{\mu\nu}\d_\beta h^{\mu\nu}
+4h_{\alpha\beta}\d_\nu h^{\beta\mu}\d_\mu h^{\alpha\nu}
\nonumber \\
+2h_{\alpha\beta}\d_\mu h^{\alpha\beta}\d^\mu h-4h_{\alpha\beta}\d_\nu h^{\alpha\mu}\d^\nu 
h_\mu^{\beta}-4h^{\mu\alpha}\d_\alpha v^\nu\d_\mu v_\nu
\nonumber \\
 -4u^\mu\d_\beta\tilde u_\nu\d_\mu h^{\nu\beta}+4\d_\nu u^\beta\d_\mu\tilde u_\beta 
h^{\mu\nu}-4\d_\nu u^\nu\d_\mu\tilde u^\beta h^{\beta\mu}+4\d_\nu u^\mu\d_\mu
\tilde u_\beta h^{\nu\beta}+
\nonumber \\
+m^2\Bigl({4\over 3}h^{\mu\nu}h_{\mu\beta}h^{\nu\beta}-
h^{\mu\beta}h_{\mu\beta}h+{1\over 6}h^3\Bigl)+4m u^\mu\tilde u^\nu\d_\mu v_\nu\Bigl\}
\nonumber \\
+{8\over\mu}\Bigl\{\epsilon^{\alpha\beta\gamma}~\Gamma^{\nu}_{\rho\alpha}~\Gamma^{\rho}_{\mu\beta}~ \Gamma^{\mu}_{\nu\gamma}
\nonumber \\
+ 3~\epsilon^{\mu\nu\rho}~\lambda_{\mu}~\Gamma_{\nu;\alpha}~(\d_{\rho}\tilde{u}^{\alpha} + \d^{\alpha}\tilde{u}_{\rho} 
- \delta^{\alpha}_{\rho}~\d_{\sigma}\tilde{u}^{\sigma})
\nonumber \\
- {3\over 2}~\epsilon_{\mu\nu\rho}~\Gamma^{\mu;\alpha}~V^{\nu}~\phi^\rho_{\,\alpha}
+ 6~m~\epsilon_{\mu\nu\rho}~\lambda^{\mu}~V^{\nu}~\tilde u^\rho+
\nonumber \\
+{3 \over 2}~\left( - m \tilde{u}^{\beta} \phi_{\beta\mu}~\lambda^{\mu} 
- {1 \over 2}~\phi^{\alpha\beta}~{\phi_{\beta}}^{\cdot\mu}~\Gamma_{\mu;\alpha}
+ m~\tilde{u}^{\beta} \phi~\lambda^{\beta} + {1 \over 2}~\phi^{\alpha\beta}~\phi~\Gamma_{\beta;\alpha}\right)
\nonumber \\
+c_{1}~\phi^{\mu\nu}~\phi_{\nu\rho}~{\phi_{\mu}}^{\rho} + c_{2}~\phi^{\mu\nu}~\phi_{\mu\nu}~\phi + c_{3}~\phi^{3}\Bigl\}.
\label{T}
\eea
This is the lowest order trilinear coupling, higher orders can be computed from higher orders
causal gauge invariance as in four dimensions \cite{10}, \cite{19}. Note that in the limit $m\to 0$
the vector-graviton field $v^\mu$ which is also contained in $V^\mu$ and $\phi_{\mu\nu}$ does
not decouple from the graviton field $h^{\mu\nu}$ in both parity-even and odd sectors. That means
the massless limit of the massive theory does not agree with the $m=0$ theory constructed without
the $v$-field. The latter is trivial as far as the physical Hilbert space is concerned whereas
the former is non-trivial.

In order to make contact with classical field theory one certainly asks: what is the classical
Lagrangian which after expansion around flat background leads to the coupling (\ref{T}). The answer
is simple as far as the $m$-independent pure $h$-terms are concerned \cite{10}: one takes the
Einstein-Hilbert Lagrangian
\be
L_{EH}=-{2\over\kappa^2}\sqrt{-g}R
\label{eh}
\ee
and expands the metric in the form
\be
\sqrt{-g}g^{\mu\nu}=\eta^{\mu\nu}+\kappa h^{\mu\nu}
\ee
using the so-called Goldberg variables. Then the terms $O(\kappa)$ agree with the $m$-independent
pure graviton terms in (\ref{T}) (if an overall factor 4 is multiplied in (\ref{eh})).
The $m$-dependent terms in the fourth line of (\ref{T}) are obtained if we add to (\ref{eh}) a
cosmological term
\be
-{2\over\kappa^2}\sqrt{-g}\,2\Lambda
\ee
and put
\be
m^2=-2\Lambda.
\label{cosmo}
\ee
The same mass value comes out from 
$
O(\kappa^0)
$ 
and 
$
O(\kappa^2)
$ 
of the expansion. Consequently, we arrive at the cosmological topologically massive gravity with a negative
cosmological constant which has been intensively studied in the literature.

The classical theory is usually expanded around anti-de-Sitter background which brings out
the behavior in the large. Our quantum theory describes the local aspects, i.e. the quantum
fluctuations around the local Minkowski space. As a consequence the two pictures are rather
different, for example, the mass of the classical 
$
AdS_3
$ 
graviton depends on the Chern-Simons coupling constant $\mu$ \cite{18} whereas we have the simple relation (\ref{cosmo}). Another difference is that the classical Chern-Simons coupling contains a quadratic term
$
\Gamma \d\Gamma
$
which does not appear in our quantum theory. Our theory is only consistent if the vector-graviton field 
$
v^\mu (x)
$ 
with the same mass $m$ as the graviton is included. This is required for quantum gauge invariance which is necessary for unitarity of the S-matrix
on the physical Hilbert space. In section 3 we have shown that the $v$-field carries physical
degrees of freedom. Therefore, it is hard to believe that the classical theory can be consistent
without the $v$-field.

\end{document}